\documentstyle[12pt]{article}
 
\begin{document}

\baselineskip=14pt plus 0.2pt minus 0.2pt
\lineskip=14pt plus 0.2pt minus 0.2pt



\begin{center}
\large{\bf  
HOLSTEIN-PRIMAKOFF/BOGOLIUBOV TRANSFORMATIONS 
AND THE MULTIBOSON SYSTEM\\}
 
\vspace{0.25in}

\large

Michael Martin Nieto\footnote{\noindent  Email: mmn@pion.lanl.gov}\\
{\it Theoretical Division, Los Alamos National Laboratory\\
University of California\\
Los Alamos, New Mexico 87545, U.S.A. \\}
and\\
{\it Abteilung f\"ur Quantenphysik \\
Universit\"at Ulm \\
D-89069 Ulm, GERMANY \\}
 
\vspace{0.25in}

 D. Rodney Truax\footnote{Email:  truax@acs.ucalgary.ca}\\
{\it Department of Chemistry\\
 University of Calgary\\
Calgary, Alberta T2N 1N4, Canada\\}
 
\normalsize

\vspace{0.3in}

{ABSTRACT}
 
\end{center}

\baselineskip=.33in

\begin{quotation}
As an aid to understanding the  {\it displacement operator}
definition of squeezed states for arbitrary systems, we investigate the
properties of systems where there is a Holstein-Primakoff or Bogoliubov 
transformation.  In these cases the  {\it ladder-operator or 
minimum-uncertainty} definitions of squeezed states are equivalent to 
an extent displacement-operator definition.   We exemplify this in
a setting where there are operators satisfying $[A, A^{\dagger}] = 1$, 
but the $A$'s are not necessarily the Fock space $a$'s; the multiboson
system.   It has been previously observed that the ground state 
of a system often can be shown to to be a coherent state.  We demonstrate 
why this must be so.  We close with a discussion of an alternative, 
effective definition of displacement-operator squeezed states.

\end{quotation}

\newpage

\baselineskip=.33in

\section{Introduction}

As has now been known and studied for some time, there are three 
equivalent,
widely-used definitions of the coherent states of the harmonic 
oscillator 
\cite{sch}-\cite{2}.  These are (1) the minimum-uncertainty, 
(2) annihilation-
(or, more generally, ladder-) operator, and (3) displacement-operator
methods.   These methods have been extended to the squeezed states of
the harmonic oscillator.  Further,  with one exception, general
coherent and squeezed states have been obtained for general systems by
these three methods.  That exception is a general definition of
squeezed states by the displacement-operator method.  

With an aim towards understanding  a general method, we can study
systems where such a definition works.  Specifically, after reviewing
the coherent and squeezed states for the harmonic oscillator and more
general systems, we focus on why displacement-operator squeezed
states often can not be obtained by a naive generalization of the
harmonic-oscillator case:  
this is when there is, in general, no Bogoliubov
transformation.   

This problem does not exist in certain systems.  In particular, we here
study the multiboson formalism of Brandt and Greenberg
\cite{brandt}, where the
multi-boson operators obey canonical commutation relations, and hence
one can proceed with calculations in the standard way.  
Elsewhere \cite{II}, we will study time-dependent systems which
have isomorphic symmetry algebras. 

We also explain the property of these various definitions of squeezed and 
coherent states which is that the ground state is a member of the set of 
coherent states.  In closing, we discuss an alternative, effective method 
for defining displacement-operator squeezed states.


\section{The Coherent and Squeezed States of the Harmonic Oscillator}

\subsection {Coherent states}

Given the canonical
commutation relations
\begin{equation}
[a,~a^{\dagger}] = 1~, ~~~~~~ [a,~ a] = 0~,  \label{coma}
\end{equation}
where we adopt the realization 
\begin{equation}
a=\frac{1}{\sqrt{2}}(x+ip),~~~a^{\dag} = \frac{1}{\sqrt{2}}(x-ip).\label{adef}
\end{equation}
The  definitions of  displacement-operator and ladder-operator coherent
states are well known. They are 
\begin{equation}
        D(\alpha)|0\rangle  = |\alpha\rangle  \label{docs}
\end{equation}
and
\begin{equation}
a|\alpha\rangle  = \alpha|\alpha\rangle~,  \label{locs}
\end{equation}
where 
\begin{equation}
D(\alpha) = \exp[\alpha a^{\dagger} - \bar{\alpha} a]
=\exp\left[-\frac{1}{2}|\alpha|^2\right]
\exp[\alpha a^{\dagger}] \exp[-\bar{\alpha} a]   \label{D}
\end{equation}
and
\begin{equation}
|\alpha\rangle = 
\exp\left[-\frac{1}{2}|\alpha|^2\right] \sum_{n} 
\frac{\alpha ^n}{\sqrt{n!}}
|n\rangle ~. 
\end{equation}
The last equality in Eq. (\ref{D}) comes from using
a Baker-Campbell-Hausdorff relation.
Observe that the definition (\ref{locs}) follows from the definition
(\ref{docs}) by
\begin{equation}
[a,~D(\alpha)] = \alpha D(\alpha)~.
\end{equation}

The coherent-state wave functions are ($m\omega/\hbar \rightarrow 1$)
\begin{equation}
\psi_{cs}(x) = \pi^{-1/4}\exp\left[-\frac{(x-x_0)^2}{2}+ip_0x\right]~, 
\label{psi}
\end{equation}
\begin{equation}
x_0 = \langle x\rangle~, 
  ~~~~ p_0 =\langle p\rangle~,
\end{equation}
\begin{equation}
Re(\alpha)=x_0/2^{1/2}~, 
\hspace{.5in} Im(\alpha)=p_0/2^{1/2}~.
\end{equation}
That is, the states are Gaussians with the width being that of the 
ground state.  

\subsection{Squeezed states}

Squeezed states \cite{ss1}-\cite{ss5}  can be
defined  by the displacement-operator method 
as the product of a unitary displacement operator and a unitary squeeze 
operator
acting on the ground state:
\begin{equation}
 D(\alpha)S(z)|0\rangle  \equiv |\alpha,z\rangle~, 
~~~~~z = z_1+iz_2 = re^{i\theta}~.
\end{equation}
$\theta$ is a phase which defines the starting time, $t_0 =
(\theta/2\omega)$.
$S(z)$ is given by
\begin{eqnarray}
S(z) & = & \exp\left[{\frac{1}{2}}za^{\dagger}a^{\dagger} -
{\frac{1}{2}}\bar{z}aa\right]  
               \label{a} \\
& = &  \exp\left[{\frac{1}{2}}e^{i\theta}(\tanh r)
a^{\dagger}a^{\dagger}\right]
\left({\frac{1}{\cosh r}}\right)^{({\small{\frac{1}{2}}+a^{\dagger}a})}
\exp\left[-{\frac{1}{2}}e^{-i\theta}(\tanh r)aa\right] \label{b} \\
& = &  \exp\left[{\frac{1}{2}}e^{i\theta}(\tanh r)
a^{\dagger}a^{\dagger}\right]
(\cosh r)^{-1/2}\sum_{n=0}^{\infty}
      \frac{({\rm sech} r -1)^n}{n!}(a^{\dagger})^n(a)^n   \nonumber \\
&~& ~~~~~~~~\times~
\exp\left[-{\frac{1}{2}}e^{-i\theta}(\tanh r)aa\right]~, \label{c}
\end{eqnarray}
where Eqs. (\ref{b})  and (\ref{c}) are
 obtained from  BCH relations.
Observe that 
\begin{equation}
D(\alpha)S(z) = S(z)D(\gamma)~, ~~~~~~
\gamma = \alpha\cosh r - \bar{\alpha} e^{i\theta}\sinh r~.
\end{equation}
Therefore, the ordering of $D$ and $S$ is only a convention.  

The squeezed-state wave functions are  given by  
a more complicated form of Eq.
(\ref{psi}).  Specifically, they are \cite{bchst}
\begin{eqnarray}
\psi_{ss} &=& D(\alpha)S(z)\psi_0 \nonumber \\
  &=& \frac{1}{\pi^{1/4}}\frac{\exp[-ix_0p_0/2]}{[{\cal S}(1 +i2\kappa)]^{1/2}}
  \exp\left[-(x-x_0)^2 \left(\frac{1}{2{\cal S}^2(1+i2\kappa)}-i\kappa\right)
    +ip_0x\right]~,
\end{eqnarray}
where
\begin{equation}
{\cal{S}} = \cosh r + \frac{z_1}{r} \sinh r
= e^r \cos^2\frac{\phi}{2} +e^{-r}\sin^2\frac{\phi}{2}~ \label{squeeze}
\end{equation} 
and
\begin{equation}
\kappa = \frac{z_2 \sinh r}{2rs}~.
\end{equation}
These wave functions are  Gaussians which, in general,  
do not
have  the width of the ground state; i.e., they are squeezed by the 
squeeze parameters ${\cal S},\kappa$.  The most commonly studied
example is when $z$ is real and positive, giving
\begin{equation}
\psi_{ss}(x) = [\pi s^2]^{-1/4}
\exp\left[-\\frac{(x-x_0)^2}{2s^2}+ip_0x\right]~, ~~~~s=e^r~.
\label{psiss}
\end{equation}

The  elements involved in $S$ actually are an SU(1,1) group 
defined by
\begin{equation}
K_+ = \frac{1}{2}a^{\dagger}a^{\dagger}~, \hspace{0.5in}
K_- = \frac{1}{2}aa~,  \hspace{0.5in} 
K_0 = \frac{1}{2}\left(N + \frac{1}{2}\right)~, \label{j}
\end{equation}
where $N=a^{\dagger}a$.  The operators $K_0, K_{\pm}$ satisfy the 
commutation relations  
\begin{equation}
[K_{0},~K_{\pm}] = {\pm}K_{\pm}~,~~~~[K_{+},~K_{-}] = -2K_{0}~.  
\label{comj}
\end{equation}
Therefore, $S$ can be given by 
\begin{eqnarray}
S(z) & = & \exp[zK_+ - \bar{z}K_-]  \label{aa} \\
     & = &  \exp[e^{i\theta}(\tanh r)K_+]
\left({\frac{1}{\cosh r}}\right)^{2K_0}
\exp[-e^{-i\theta}(\tanh r)K_-]~.  \label{bb}
\end {eqnarray}
The commutation relations (\ref{coma}) and (\ref{comj}) close with
\[
[K_{+}, ~a^{\dagger}]=0~,~~~~~~ [K_-,~a^{\dagger}] = a~, ~~~~~~
[K_{+}, ~a]  =  -~a^{\dagger}~,  
\]
\begin{equation}
[K_{-},~a]=0~,  ~~~~~~
[K_{0}, ~a^{\dagger}]=
\mbox{$1\over2$}~a^{\dagger}~,
~~~~~~[K_{0},a]=-\mbox{$1\over2$}~a.
\end{equation}

The ladder-operator definition of the squeezed states is  
\begin{equation}
[\mu a - \nu a^{\dagger}]|\alpha,z\rangle = \beta|\alpha,z\rangle~.
\end{equation}
Again this follows from the displacement-operator definition 
because
\begin{eqnarray}
b \equiv   S(z)^{-1}aS(z) &=& 
(\cosh r)a +e^{i\theta}(\sinh r) a^{\dagger}~,    \nonumber  \\
b^{\dagger} \equiv S(z)^{-1}aS(z) &=& 
(\cosh r)a^{\dagger} +e^{-i\theta}(\sinh r) a~.    \label{btrans}
\end{eqnarray}
where
\begin{equation}
[b,~b^{\dagger}]=1~, ~~~~~~b\equiv\mu a +\nu a^{\dagger}~, ~~~~~~
        |\mu|^2 - |\nu|^2 = 1~.
\end{equation}
Eq. (\ref{btrans}) is a Holstein-Primakoff \cite{hp} or 
Bogoliubov \cite{bog} transformation.  When such a transformation
exits, such as for the harmonic oscillator and for some other cases
\cite{bt1}-\cite{bt3},
there is no problem defining  displacement-operator squeezed states. 
However, such a transformation does not always exist, and that is at
the crux of the problem of finding a general definition for
displacement-operator squeezed states.

Lastly, we note the time-dependent uncertainties in $x$ and $p$.  They
are \cite{uncert}
\begin{equation}
[\Delta x(t)]^2_{(\alpha,z)}
 = \mbox{$1\over2$}\left[s^2\cos^2\omega t
 ~+~\frac{1}{s^2}\sin^2\omega t\right]~, 
\end{equation}
\begin{equation}
[\Delta p(t)]^2_{(\alpha,z)}
 = \mbox{$1\over2$}\left[\frac{1}{s^2}\cos^2\omega t~
+~s^2\sin^2\omega t\right]~, 
\end{equation}
\begin{equation}
[\Delta x(t)]^2_{(\alpha,z)}[\Delta p(t)]^2_{(\alpha,z)}
=\frac{1}{4}\left[ 1~+~\frac{1}{4}\left(s^2-\frac{1}{s^2}\right)^2
         \sin^2[\omega t]\right]~.
\end{equation}


\section{Generalized Coherent and Squeezed States}

As discussed in Ref. \cite{nt}, generalizations of the
displacement-operator and ladder-operator coherent states have been
widely discussed and studied \cite{jb,gcs1,gcs2,gcs3}. Also,
a generalization of the minimum-uncertainty coherent states  was 
found \cite{n1,n2},
and this method turned out to also yield the generalized squeezed states
as a byproduct.  

Recently, we gave a generalized ladder operator method to define
squeezed states for arbitrary systems \cite{nt}, and there we pointed
out the problem which is at the crux of the present study.  In general
there is no Bogoliubov transformation and hence no connection between 
the ladder-operator and displacement-operator methods for defining
squeezed states.  

This can be exemplified by considering the ordinary squeeze operator
acting on the ground state, with no displacement operator:
\begin{equation}
S(z)|0\rangle = |z\rangle  ~.
\end{equation}
In this form, $S(z)$ is the SU(1,1) displacement operator, and hence
the states $|z\rangle$ are the SU(1,1) coherent states. 
Note that these coherent states have only even occupation numbers  in
the number basis.  (Indeed,
recall that one of the early names for the squeezed states was
``two-photon coherent states" \cite{ss1}.)  

But if $S$ is the
displacement operator for SU(1,1), what is the SU(1,1) squeeze
operator?  A first guess would be to square the elements of $S$,
i.e., to square $aa$ and $a^{\dagger}a^{\dagger}$ to yield operators
exponentiated to the fourth power.   
But this leads to operators that are
not well-defined \cite{us1,us2};  that is,  the operators
\begin{equation}
U_j = exp[\hat{z}_j(a^{\dagger})^j - \hat{\bar{z}}_j(a)^j]~, 
~~~~j = 3,4,5,\dots .
\end{equation}
So, there is no naive higher-order squeezing. 
Another way to state this is that there exist no simple operators which
obey   
\begin{equation}
\hat{S}(y)^{-1} aa \hat{S}(y) = \mu aa + \nu a^{\dagger}a^{\dagger}~.
\end{equation}
That is, there is no Bogoliubov transformation for the SU(1,1)
elements.   Hence, there is no obvious way to define the SU(1,1)
squeezed states by the displacement-operator method.  


\section{Multiboson Operators}

In a program to circumvent the problems with naive multiboson squeezing,
a productive collaboration \cite{d1}-\cite{d6} proposed using the
generalized Bose operators of Brandt and Greenberg \cite{brandt}.
These latter two observed that if one defines the operators 
\begin{equation}
A_j = \sum_{k=0}^{\infty} \alpha_{jk} (a^{\dagger})^ka^{k+j}~,
~~~~j \ge 2~, 
\end{equation}
\begin{equation}
 \alpha_{jk} = \sum_{l=0}^k\frac{(-1)^{k-l}}{(k-l)!}
\left[\frac{1+[[l/j]]}{l!(l+j)!}\right]^{1/2} e^{i\rho_l}~,
\end{equation}
where we denote the greatest-integer function by $[[y]]$, and the 
${\rho_l}$ are arbitrary phases.  Then, we have 
\begin{equation}
[A_j^{\dagger},~A_j] = 1~.   \label{Aj}
\end{equation}
That is, these functions satisfy the canonical commutation relations
even though they are not the ordinary boson operators.  They also 
satisfy
\begin{equation}
[N,A_j] = [a^{\dagger}a,A_j]= -jA_j~,
\end{equation}
and 
\begin{eqnarray}
A_j|jn + k\rangle &=& \sqrt{n}|j(n-1) + k\rangle~, \\
A_j^{\dagger}|jn+k\rangle &=& \sqrt{(n+1)}|j(n+1) + k\rangle~,
~~~~~~0\leq k < j~.
\end{eqnarray}
Note that for a given $j$ we have $j$ different  sets of  states.
Each of them starts at a different  lowest state $|k\rangle$, 
where $0 \leq k <
j$; i.e., $|0\rangle$, $|1\rangle$, $|2\rangle$,
\dots $|j-1\rangle$.

If one acts  on  eigenstates of $N$, then 
from the normal-ordering theorems of Wilcox \cite{wilcox}, 
a very useful 
form of $A_j$ can be given   \cite{r}
\begin{equation}
A_j^{\dagger} = 
\left[[[\tilde{N}/j]]\frac{(\tilde{N}-j)!}{\tilde{N}!}\right]^{1/2}
(a^{\dagger})^j~,
\end{equation}
where $\tilde{N}$ is the eigenvalue of the operator $N$ in the number operator basis.

The collaboration of Refs. \cite{d1}-\cite{d6} concentrated on  investigating
the properties of the states defined by 
\begin{equation}
D(\alpha)V(z) |0\rangle = 
D(\alpha)\exp[z A_j^{\dagger} - \bar{z} A_j] |0\rangle = 
|\alpha, z_j\rangle~.  \label{them}
\end{equation}
In other words, they took an ordinary coherent state and then 
squeezed this state by the j-photon operators of $A_j$ and $A_j^{\dagger}$.
(Also, they studied \cite{d5} the properties of  states obtained from
a generalized  set of Weyl-Heisenberg operators, $A_j^{\eta}$.)


\section{Coherent and Squeezed States for the Multiboson
Systems}

\subsection{Coherent states}

Now, from our point of view, of finding general and consistent
methods of obtaining coherent and squeezed states, 
another path suggests itself.  Since the $A_j$'s obey the canonical
commutation relations of Eq. (\ref{Aj}), which are identical in form
to  Eq.  (\ref{coma}),   this means one can use {\it these}
operators in displacement operators.  That is, we consider the 
operator $V$ of equation (\ref{them}) not to be a multiboson
squeeze of a coherent state, but rather a multiboson displacement 
operator: 
\begin{equation}
D_j(\alpha) = \exp[\alpha A_j^{\dagger} - \bar{\alpha} A_j]
=\exp\left[-\frac{1}{2}|\alpha|^2\right]
\exp[\alpha A_j^{\dagger}] \exp[-\bar{\alpha} A_j]  ~.
\end{equation}
Therefore, the multi-boson coherent states are 
\begin{equation}
|\alpha (j,k)\rangle = D_j(\alpha)|k\rangle
= \exp\left[-\frac{1}{2}|\alpha|^2\right]
\sum_{n=0}^{\infty}\frac{\alpha^n}{\sqrt{n!}} |jn+k\rangle~.
\end{equation}
Again observe 
 that for a given $j$ we have $j$ different  sets of (coherent) states.
Each of them again starts 
 at a different  lowest state $|k\rangle$, where $0 \leq k <
j$; i.e., $|0\rangle$, $|1\rangle$, $|2\rangle$,
\dots $|j-1\rangle$. That is why we label the states
 by the couple $(j,k)$.  [The states
$|\alpha(j,0)\rangle$ were studied in Ref. \cite{d4}.]

These coherent states are, of course, consistent with the ladder-operator
definition,
\begin{equation}
A_j |\alpha (j,k)\rangle = \alpha |\alpha (j,k)\rangle ~.
\end{equation}
By using the number-state basis of the wave functions,
\begin{equation}
\psi_n = \left(\frac{a_o}{\pi^{1/2} 2^n n!}\right)^{1/2}
\exp\left[-\frac{1}{2}a_0^2x^2\right]
H_n(a_0x)~, 
\end{equation}
where $a_0^2= (m\omega/\hbar)$ will 
now be set to $1$ and the $H$ are the Hermite
polynomials, one can write the normalized coherent state wave functions as 
\begin{equation}
\psi_{cs}(j,k)(x) = \pi^{-1/4}
\exp\left[-\frac{1}{2}\left(|\alpha|^2 + x^2\right)\right] 
I_{(j,k)}(\alpha,x)~,
\end{equation}
where $I$ is the sum
\begin{equation}
I_{(j,k)}(\alpha,x)= \sum_{n=0}^{\infty} 
\frac{\alpha^n H_{jn+k}(x)}{[n!(jn+k)!2^{jn+k}]^{1/2}} ~.
\end{equation}
Note that for  $(j,k)= (1,0)$, we obtain the usual 
generating function \cite{genf} for the ordinary coherent states
result, 
\begin{equation}
I_{(1,0)}(x) = \exp[\sqrt{2}\alpha x -\alpha^2/2]~.
\end{equation}

The ``natural quantum operators" for this system are \cite{n1,n2}
 (in dimensionless units)
\begin{equation}
X_j \equiv \frac{1}{\sqrt{2}}[A_j + A_j^{\dagger}]~,  ~~~~~~~
P_j \equiv \frac{1}{i\sqrt{2}}[A_j - A_j^{\dagger}]~,
\end{equation}
But then, the Heisenberg-Weyl algebra tells us immediately that these
are the operators directly connected to the minimum-uncertainty method.   
Therefore, we have that \cite{nt}
\begin{equation}
(\Delta X_j)^2_{(j,k)}= 1/2~,~~~~(\Delta P_j)^2_{(j,k)} = 1/2~.
\end{equation}

We can also obtain information for the uncertainties of the physical position
and momentum, $x$ and $p$.  We immediately observe that 
\begin{equation}
\langle x \rangle_{(j,k)} = \langle p \rangle_{(j,k)} = 0~, ~~~~j>1~.
\end{equation}
(For j=1 we have the ordinary harmonic oscillator).  
For $j>2$, we have, then, that 
\begin{eqnarray}
\langle x^2 \rangle_{(j,k)}
&=&
(\Delta x)^2_{(j,k)}
=\langle p^2 \rangle_{(j,k)}= (\Delta p)^2_{(j,k)}  \nonumber \\
 &=& 
\exp[-|\alpha|^2] \sum_{n=0}^{\infty}\frac{|\alpha|^{2n}}{n!}
 [jn+k+\mbox{$1\over2$}]~
  \nonumber \\ 
&=& 
\mbox{$1\over2$} + k +j|\alpha|^2~, ~~~~~~j>2~.
\end{eqnarray}

The case $j=2$ is slightly more complicated because the operators $x^2$
and $p^2$ connect different number states in the expectation values.
In particular, 
\begin{eqnarray}
\langle x^2 \rangle_{(2,k)}&=&(\Delta x)^2_{(2,k)}
=\mbox{$1\over2$} + k +2|\alpha|^2 + C_{(2,k)}  \\
\langle p^2 \rangle_{(2,k)}&=&(\Delta p)^2_{(2,k)}
=\mbox{$1\over2$} + k +2|\alpha|^2 - C_{(2,k)} ~,
\end{eqnarray}
where 
\begin{equation}
 C_{(2,k)}=\mbox{$1\over2$}[\langle a^2 \rangle_{(2,k)}
+\langle (a^{\dagger})^2 \rangle_{(2,k)}]~, 
\end{equation}
which evaluates to
\begin{equation}
 C_{(2,k)}=(\alpha + \bar{\alpha})\exp[-|\alpha|^2]
\sum_{n=0}^{\infty}\frac{|\alpha|^{2n}}{n!}
\left[\frac{(n+1+k/2)(n+1/2+k/2)}{n+1}\right]^{1/2}~.
\end{equation}

\subsection{Squeezed states}

Because the $A_j$'s define a Heisenberg-Weyl algebra, one can therefore
define an SU(1,1) squeeze algebra in the normal way:

\begin{equation}
K_{j+} = \frac{1}{2}A_j^{\dagger}A_j^{\dagger}~, \hspace{0.5in}
K_{j-} = \frac{1}{2}A_jA_j~,  \hspace{0.5in} 
K_{j0} = \frac{1}{2}\left(A_j^{\dagger}A_j + \frac{1}{2}\right)~.
 \label{jj}
\end{equation}
Then all these $A_j$'s and $K_j$'s again have the same commutation
relations as before, and so all the results of the ordinary harmonic
oscillator coherent and squeezed states goes through in the same
manner, only with the $a$'s being changed into the $A_j$'s. 
The squeeze operators are therefore
\begin{eqnarray}
S_j(z) & = & \exp[zK_{j+} - \bar{z}K_{j-}]  \label{jaa} \\
     & = &  \exp[e^{i\theta}(\tanh r)K_{j+}]
\left({\frac{1}{\cosh r}}\right)^{2K_{j0}}
\exp[-e^{-i\theta}(\tanh r)K_{j-}]~,  \label{jbb}
\end {eqnarray}
where 
\begin{equation}
z = r e^{i\theta}~,
\end{equation}
meaning the squeezed states are 
\begin{equation}
D_j(\alpha)S_j(z)|k\rangle = |\alpha, z(j,k)\rangle~.
\end{equation}

Furthermore, all the mathematics of the ordinary squeezed states  
follows automatically, just changing notation.  For example, there is a
Bogoliubov transformation:
\begin{eqnarray}
B_j \equiv   S_j(z)^{-1}A_jS_j(z) &=& 
(\cosh r)A_j +e^{i\theta}(\sinh r) A_j^{\dagger}~,    
\label{btransj}  \\ 
B_j^{\dagger} \equiv S_j(z)^{-1}A_jS_j(z) &=& 
(\cosh r)A_j^{\dagger} +e^{-i\theta}(\sinh r) A_j~.   
 \label{bdtransj}
\end{eqnarray}
where
\begin{equation}
[B_j,~B_j^{\dagger}]=1~, ~~~~~~B_j\equiv\mu A_j +\nu A_j^{\dagger}~, 
~~~~~~    |\mu|^2 - |\nu|^2 = 1~.
\end{equation}
This means, of course, that there is an equivalent ladder-operator
definition of these squeezed states:
\begin{equation}
[\mu A_j - \nu A_j^{\dagger}]|\alpha,z(j,k)\rangle 
= \beta|\alpha,z(j,k)\rangle~.
\end{equation}

Again, from the the Heisenberg-Weyl algebra, it follows that 
\begin{equation}
(\Delta X_j)^2_{ss}(\Delta P_j)^2_{ss} = 1/4~.
\end{equation}
Of course, being squeezed states the above equality holds at $t=0$
and oscillates, and the uncertainty in each quadrature also oscillates.


\section{The Ground State as a Coherent State}

In finding the coherent and squeezed states for general systems, 
it has been noted 
that the ground state (or extremal state) is always a member of 
the set of coherent states \cite{nt,n1}.  This is also true in 
the multi-boson case and we want to show that why it is true in 
general.  Before continuing, however, note that this makes 
intuitive sense.  The ground state is the closest quantum state 
to zero motion, which corresponds to a classical particle at 
rest.  Therefore, the most-classical like states should include 
this state.

Starting from a minimum-uncertainty Hamiltonian system, the 
classical Hamiltonian
is transformed to classical variables that vary as the sin and the 
cosine of the classical $\omega t$.  In these variables the 
Hamiltonian can be written as 
\begin{equation}
H_{cl} = X^2/2 + P^2/2~.
\end{equation}
This is harmonic-oscillator like.  Indeed, for the rest of this 
discussion keep the harmonic oscillator in mind for intuitive aid. 

When the classical variables are changed to quantum operators, it is 
found that 
\begin{equation}
X = \frac{{\cal A} + {\cal A^{\dagger}}}{\sqrt2}~, ~~~~~
P = \frac{{\cal A} - {\cal A^{\dagger}}}{i\sqrt2}~,
\end{equation}
where the ${\cal A}$'s are the lowering operators of the system.  
In general, these operators may be $n$-dependent or have to be 
made Hermitian with respect to the adjoint, but the statement holds.

Therefore, the states which minimize the uncertainty relation between
$X$ and $P$,
\begin{equation}
[X,P] = iG~,  \label{com}
\end{equation}
are those (squeezed) states which satisfy the eigenvalue equation
\begin{equation}
\left[X +\frac{i\Delta X}{\Delta P}P\right]\psi_{mus} =
\left[\langle X\rangle +
\frac{i\Delta X}{\Delta P}\langle P\rangle \right]\psi_{mus} ~.
\label{mus}
\end{equation}
[When dealing with symmetry, non-Hamiltonian systems, the starting 
point for the study is here, simply considering the implications
of the commutation relation (\ref{com}).]

These states are, in general, squeezed states.  This can be seen 
by writing $X$ and $P$ in terms of ${\cal A}$ and 
${\cal A}^{\dagger}$.  Then the left hand side of Eq. (\ref{mus})
is proportional to a linear combination of ${\cal A}$ and 
${\cal A}^{\dagger}$, just like after any Bogoliubov transformation.  
To change to a coherent state, the relative uncertainties must be 
equal, i.e., $(\Delta X)/(\Delta P)= 1$.  but then the left hand side
of Eq. (\ref{mus}) is proportional simply to ${\cal A}$.  Then taking 
the case corresponding to the smallest classical motion, 
$\langle X\rangle=\langle P\rangle = 0$, one is left with the equation
\begin{equation}
{\cal A}\psi_{mus} = 0~.
\end{equation}
But the state that is annihilated by the lowering operator is the 
ground state. 


\section{An Alternative, Effective Definition for Displacement-Operator
Squeezed States}

We close with a comment on how an alternative method can be used 
to define "displacement operator" squeezed states.
This method can be used for the systems under discussion:  systems with 
minimum-uncertainty, ladder-operator, and displacement operator 
coherent states, but only minimum-uncertainty or ladder-operator
squeezed states.  An example, where it has been used, suffices 
to explain the procedure.  

The even and odd coherent states \cite{eo, eocs} 
are defined in terms of the double
destruction operator:
\begin{equation}
aa|\alpha\rangle_{\pm} = \alpha^2|\alpha\rangle_{\pm}~.
\end{equation}
They also can be defined in terms of an unusual displacement 
operator, 
\begin{equation}
|\alpha\rangle_{\pm}= 
D_{\pm}(\alpha)|0\rangle
=\left[2(1\pm \exp[-2|\alpha|^2]\right]^{-1/2}
          \left[D(\alpha) \pm D(-\alpha)\right] |0\rangle~,
\end{equation}
where $D$ is the ordinary coherent state displacement operator.

The even and odd squeezed states are generalized to those 
states satisfying the eigenvalue equation
\begin{equation}
\left[\left(\frac{1+q}{2}\right) aa
+\left(\frac{1-q}{2}\right) a^{\dagger} a^{\dagger} \right]\psi_{ss} 
=\alpha^2 \psi_{ss} .   \label{Kss}
\end{equation}
The solutions are  \cite{nt}
\begin{equation}
\psi_{Ess}=N_E\exp{\left[-\frac{x^2}{2}(q+\sqrt{q^2-1})\right]}
\Phi\left(\left[\frac{1}{4}+\frac{\alpha^2}{2\sqrt{q^2-1}}\right],~
\frac{1}{2};~x^2\sqrt{q^2-1}\right)~,
\end{equation}
\begin{equation}
\psi_{Oss}=N_O ~x\exp{\left[\frac{-x^2}{2}(q+\sqrt{q^2-1})\right]}
\Phi\left(\left[\frac{3}{4}+\frac{\alpha^2}{2\sqrt{q^2-1}}\right],~
\frac{3}{2};~x^2\sqrt{q^2-1}\right)~,
\end{equation}
where  $\Phi(a,b;c)$ is the confluent hypergeometric function
$ \sum_{n=0}^{\infty} \frac{(a)_n c^n}{(b)_n~n!}$.
In the limit $q\rightarrow 1$, these become the even and odd coherent states.

But there are no displacement operator squeezed states because there does 
not exist a  unitary (Bogoliubov) transformation that can rotate
$aa$ into a linear combination of $aa$ and $a^{\dagger}a^{\dagger}$.  
Therefore, an alternative idea is to simply use the 
ordinary squeeze operator, $S$, with the given 
displacement operator, and call these the 
displacement-operator squeezed states \cite{kss}.  Here that 
would be 
\begin{equation}
D_{\pm}(\alpha)S(z)|0\rangle = |\alpha,z\rangle_{\pm}~.
\end{equation}

In Ref. \cite{eoss} these minimum-uncertainty and ``displacement-operator"
were compared and found to be similar in nature.  Since the 
``displacement-operator" states are more amenable to analytic calculations, 
they were then used for exploration of the time-dependence of the 
even and odd states of a trapped ion.

This, then, is a viable alternative to mathematically rigorous
siplacment-operator squeezed states.


\section*{Acknowledgements}

We wish to thank G. M. D'Ariano and Wolfgang Schleich for 
helpful conversations on these topics.  
MMN acknowledges the support of the United States Department of 
Energy and the Alexander von Humboldt Foundation.  DRT acknowledges
a grant from the Natural Sciences and Engineering Research Council 
of Canada.


\end{document}